\documentclass[%
superscriptaddress,
preprint,
nofootinbib,
nobibnotes,
 amsmath,amssymb,
pra,
longbibliography %Added by Igor
[]{revtex4-1}

\usepackage[utf8]{inputenc}
\usepackage[usenames,dvipsnames]{xcolor}
\usepackage[colorlinks,citecolor=BlueViolet,linkcolor=RedOrange,urlcolor=BlueViolet]{hyperref}
\usepackage{rsfso}

\usepackage{amsmath,color}
\usepackage{graphicx}% Include figure files
\usepackage{dcolumn}% Align table columns on decimal point
\usepackage{bm}
\usepackage{amsfonts}
\usepackage{amsthm}
\usepackage{comment}
\usepackage{ulem} %added by Igor
\usepackage{breqn}%break equations over 2 lines
\newcommand{\Zim}{Z^{\prime\prime}}
\newcommand{\Keff}{K_{\textrm{eff}}}

\newcommand{\LB}[1]{{\color{black} {#1}}}
\newcommand{\LBB}[1]{{\color{black} {#1}}}
\newcommand{\ML}[1]{{\color{black} {#1}}}
\newcommand{\MLR}[1]{{\color{black} {#1}}}

\begin{document}
\title{Drastic reduction of dynamic liquid-solid friction \\ in supercooled glycerol}
\title{Collective solid state modes control friction of supercooled glycerol}
\title{Collective excitations in the solid control friction of supercooled glycerol}
\title{Hydrodynamic friction of supercooled glycerol on excitable surfaces}
\title{Anomalous friction of supercooled glycerol on mica}

\author{Mathieu Lizée}\email{mathieu.lizee@ens.fr, lyderic.bocquet@ens.fr}	
 \author{Baptiste Coquinot}	
\author{Guilhem Mariette}
 \author{Alessandro Siria}
 \author{Lydéric Bocquet,$^*$\\
  	\normalsize{$^{1}$Laboratoire de Physique de l'Ecole Normale Sup\'erieure, ENS, Universit\'e PSL}, \\
	\normalsize{CNRS, Sorbonne Universit\'e, Universit\'e de Paris, 75005 Paris, France}\\
}
% \affiliation{Laboratoire de Physique de l'Ecole Normale Sup\'erieure, ENS, Université PSL, CNRS, Sorbonne Université, Université Paris-Cité, 75005 Paris France}
 
%\affiliation{Laboratoire de Physique de l'Ecole Normale Sup\'erieure \\ ENS,  Universit\'e PSL}

\begin{abstract}
 %As low energy electronic modes are believed to control liquid friction in carbon nanotubes, a fine understanding of dynamical aspects of slippage is now required. 
% \LB{A reprendre format Nature Com ou Physics}
 \LB{The fundamental understanding of friction of liquids on solid surfaces remains one of the key knowledge gap in the transport of fluids. While the standard perspective emphasizes on  solid-liquid commensurability and wettability effects as key ingredients, recent works have highlighted the crucial role of the internal dynamics of the solid in the liquid-solid dissipation, in the form of mechanical or electronic excitations.}
%This study addresses the influence of internal liquid dynamics on liquid-solid friction. 
\LB{Here, we take advantage of}
the wide range of relaxation timescales in a supercooled liquid \LBB{to explore how surface friction depends on the fluid molecular dynamics.} We use a \LB{dedicated} tuning-fork-based AFM to measure the hydrodynamic slippage of supercooled glycerol on mica \LBB{as a function of temperature}. % at 30 kHz. 
\LBB{In contrast to expectations and recent molecular dynamics predictions, our experiments} report a 2-order of magnitude increase of \LB{the slip length} with decreasing temperature by only 30$^\circ$C. \LB{Even more counterintuitively, we demonstrate that the liquid-solid friction coefficient %decreases upon cooling, and 
exhibits an anomalous, non-arrhrenian dependence on temperature, at odd with standard predictions based on activated dynamics. }
\LB{While several possible physical mechanisms are investigated, we show that this unconventional friction behavior is consistent with a dynamical contribution 
rooted in the contribution of internal mechanical excitations of the solid to the fluid-solid interfacial dissipation. This behavior opens new perspectives to control hydrodynamic flows by properly engineering the solid internal excitations.}
%More importantly, as the bulk liquid dynamics are slowed with decreasing temperature, we report a sharp drop of the interfacial friction coefficient in contrast with the usual assumption of thermally activated interfacial dynamics. To rationalize this original behavior, we account for the contribution of solid fluctuations to liquid friction. We show that a minimalistic single phonon-branch model of the mica surface yields semi-quantitative agreement with our measurements. In this picture, the liquid's relaxation rate is the tuning knob between two friction regimes where the wall is seen either as a static corrugated potential or as a thermally fluctuating surface. Remarkably, this study bridges soft and hard condensed matter: hydrodynamic flow controlled by the solid's dynamical modes.
\end{abstract}

%Autre proposition d'abstract
\begin{abstract}
	{The fundamental understanding of  friction of liquids on solid surfaces remains one of the key knowledge gaps in the transport of fluids. While the standard perspective emphasizes the role of wettability and commensurability, recent works have unveiled the crucial role of the solid's internal excitations, whether electronic or phononic, on liquid-solid dissipation. In this work, we take advantage of the considerable variation of the molecular timescales of supercooled glycerol under mild change of temperature, in order to explore how friction depends on the liquid's molecular dynamics. Using a dedicated tuning-fork-based AFM to measure the hydrodynamic slippage of glycerol on mica, we report a 2-order of magnitude increase of the slip length with decreasing temperature by only 30$^\circ$C. However the solid-liquid friction coefficient is found to be a non monotonous function of the fluid molecular relaxation rate, $f_\alpha$, at odd with an expected arrhenius behavior. 
%While the  decrease of friction with the liquid molecular rate -- observed at low temperature - is in agreement with predictions, 
In particular, the linear increase of friction with the liquid molecular rate measured at high temperature cannot be accounted for by existing modelling. 
%This non-ahrrenian behavior is at odd with theoretical expectations of a inverse dependency with $f_\alpha$. %the liquid relaxation rate. 
	%and recent molecular dynamics predictions, which the decrease of friction with 
%	As glycerol modes are blue-shifted, friction first decreases to reach a minimum at 34 MHz in complete agreement with theoretical predictions for a static wall. Above 34 MHz however, further stiffening of glycerol modes results in a linear increase of friction with the mode's frequency, in strong contrast with the standard Arrhenius picture. 
We show that this unconventional and non-arrhenian friction is consistent with a contribution of the solid's phonons to the liquid-solid friction. %In this picture, the 34 MHz minimum corresponds to a minimal overlap of liquid and solid modes. 
This dynamical friction opens new perspectives to control hydrodynamic flows by properly engineering phononic and electronic excitation spectra in channel walls.}
\end{abstract}

\maketitle

%A citer \cite{servantie2008temperature,long2019superlubricity,qin2022atomistic,xiang2023observation}

%The fundamental understanding of friction of liquids on solid surfaces remains one of the key knowledge gap in the transport of fluids. \cite{Faucher2019}.
Solid-liquid friction is usually accounted for by the hydrodynamic slip length, defined as {the} ratio between the bulk viscosity of the fluid and the solid-liquid friction coefficient. The topic has been the subject of \MLR{many} investigations \cite{bocquet_nanofluidics_2010,aluru2023fluids} based on a considerable body of experimental and theoretical work and it is now established that interfacial friction is determined to a large extent by surface energy, {with slippage being} promoted by hydrophobicity \cite{bocquet_nanofluidics_2010}. However, this picture is incomplete and fails to explain some experimental \MLR{measurements} on liquid-solid friction. 
%Coupling slip-enhanced liquid flow to ion transport allows to design mechano-sensitive systems capable to probe nanoscale pressure gradients \cite{marcotte2020mechanically}. 
%Unfortunately, water slippage remains elusive and has only been measured on a handful of very specific systems including atomically smooth graphite \cite{maali_measurement_2008} or the inner walls of carbon nanotubes. On the fundamental side, slippage is the ratio of bulk viscosity and interfacial friction. While the first is well defined and keeps its bulk value down to the nanoscale\cite{bocquet_nanofluidics_2010}, the latter relies on out-of-equilibrium couplings between crystalline solids and mesoscopic liquid flows and is notoriously hard to measure. 
%Based on a considerable body of experimental work, it has been well established that slippage is determined to a large extent by surface energy and thus promoted by hydrophobicity \cite{bocquet_nanofluidics_2010,huang_water_2008}. 
% The slip length, usually defined as the distance where the velocity profile extrapolates to zero inside the solid, corresponds to the ratio of viscosity to liquid-solid friction. The study of liquid-solid friction has established  as a central topic in nanofluidics, with the promise that low-friction materials could allow high permeability-high selectivity membranes for water filtration  
%This interpretation 
%has nonetheless 
%{ It has notably} failed to 
\MLR{In particular, it is} insufficient to account for the radius-dependent and ultra-low friction observed in carbon nanotubes and graphene-based nanochannels \cite{Holt2006,secchi_massive_2016,Tunuguntla2017,xie_fast_2018}, as well as the difference of friction inside nanotubes versus flat graphite \cite{maali_measurement_2008}. Recently, and beyond the picture of an inert wall, the role of {solid dynamics} on interfacial friction was put forward. 
%
%While several experiments have shown a strong influence of adsorbed liquids on the band structure and lifetime of phonons in the underlying solid \cite{warburton2023identification, babaei2016mechanisms}, 
Numerical simulations revealed indeed that mechanical fluctuations of the confining wall do affect wetting, slippage and diffusive transport inside the liquid \cite{ma2015water,marbach2018transport,noh2022effect}, while, conversly, several experiments have shown a strong influence of adsorbed liquids on the band structure and lifetime of phonons in the underlying solid \cite{warburton2023identification, babaei2016mechanisms}.
%In recent years, there has been increasing evidence from numerical simulations that fluctuations in the solid state play a central role in hydrodynamic friction 
But even more counterintuitively, the couplings of \MLR{electronic -- plasmon-like -- excitations in metals with the collective charge fluctuations in liquids was shown to generate strong fluctuation-induced contributions to friction} \cite{kavokine_fluctuation-induced_2022,bui2023classical}. This electronic friction explains the peculiar radius dependency of liquid slippage in \MLR{carbon} nanotubes \MLR{\cite{secchi_massive_2016,kavokine_fluctuation-induced_2022}} and the anomalous electron cooling of graphene in contact with water \cite{yu2023electron}.
%Coulombic coupling between solid-state electronic modes and charge fluctuations in the liquid account for the peculiar radius dependency of liquid slippage in nanotubes \cite{kavokine_fluctuation-induced_2022,bui2023classical} and for electronic cooling in graphene put in contact with water \cite{yu2023electron}. 
%Such couplings of hydrodynamic flows with the internal dynamics of solids were also invoked to explain
{Reversly, it was reported that liquid friction can induce electronic currents in graphene. This measurement demonstrates a momentum transfer from liquid flows to electrons which is believed to be mediated by acoustic phonons \cite{lizee_strong_2023,kral2001nanotube,coquinot_quantum_2023}.}  %Despite these first hints, the role of \MLR{solid state excitations -- either phononic or electronic -- }on liquid-solid friction has never been measured directly and remains elusive.
%the dependency of \LB{liquid-solid friction on the solid and liquid fluctuation} spectra remains \LB{elusive}.

%new channels of dissipation were unveiled, via
Interestingly, the emergence of couplings between the liquid fluctuations and the solid (electronic or phononic) excitations is reminescent of the longstanding quest for electronic and phononic contributions to solid-on-solid friction. Recently, these dissipation channels were elegantly disentangled by showing a strong drop of non-contact friction at the superconducting transition of Nb thin films \cite{kisiel2011suppression}. 
%While electronic friction is suppressed by the superconducting gap, the distance dependence of the remaining contribution shows a striking agreement with a phononic decay mechanism. 
Independently, solid-on-solid friction of mica on graphene was shown to be enhanced by an intercalated water layer \cite{lee2017enhancement}, via a broadening of the phonon spectrum of graphene by water.   
%The demonstration that intercalated layers of water between mica and graphene can increase solid friction then provided the first bridge to the world of liquids \cite{lee2017enhancement}. Using density functional theory calculations, the authors explained that water broadens the phonon spectrum of graphene, thus opening new channels for momentum and energy transfer. 
How these concepts extend to liquid-solid friction remains largely unknown, and despite first hints, the role of \MLR{solid state excitations -- either phononic or electronic -- }on liquid-solid friction has never been measured directly and remains elusive.
%In complete analogy with solid friction, it is natural that liquid friction shall depend on solid state excitations as well. 
Such a tunable picture of the interface opens countless possibilities of nanoscale flow engineering, via the controlled electronic properties of channel's wall \cite{coquinot_quantum_2023}.
%in ultra high vacuum down to 5.8K. Friction \textit{versus} As the sample is heated towards its superconducting critical temperature ($T_c=9.2 K$) and its BCS gap closes, non-contact friction rises rapidly up to its metallic value at which it saturates for $T>T_c$. 

In this study, we address this question from a reversed perspective: instead of studying liquid friction on a variety of solids with different phononic or electronic properties, we rather vary the excitation spectrum of the liquid. 
To this end, we leverage the dynamical slowdown of supercooled liquids whose molecular timescale typically follows the Vogel-Fulcher-Tammann law: $\tau \propto \textrm{exp}(E_a/k_B(T-T_g))$ above their glass transition temperature $T_g$ (here $E_a$ is an activation energy -- see SI Section I for details). For convenience, we use pure glycerol, which is supercooled close to room temperature. Cooling by a few tens of Kelvins causes a drastic red-shift of glycerol's spectral modes, as measured previously in mechanical and dielectric spectroscopy \cite{lunkenheimer_glassy_2000,davies1973viscoelastic,harrison1976dynamic}. 
How this frequency shift in the liquid molecular dynamics affects liquid-solid friction is the main objective of this experimental study. 
%We then measure its friction on a well-controlled solid surface and hope to disentangle bulk dynamical properties from the specific surface mechanism which control friction.}
%

\LB{Accordingly, we investigate the surface dynamics of supercooled glycerol on mica surfaces, in a range of temperature $0-35^\circ$C, corresponding to a 10\% change in absolute temperature.} 
%The internal relaxation rate is tuned in-situ and reversibly by controlling temperature. As a model solid surface, 
The cleaved (001) surface of muscovite mica has a honeycomb structure, with a molecularly smooth corrugation \cite{khan2010digitally}. This atomic smoothness makes mica a \MLR{sample} of choice for surface force investigations of liquid interfaces, evidencing \textit{e.g.} a \MLR{signature} molecular structuration \MLR{of} the liquid. Interestingly, \MLR{a no-slip condition (corresponding to an immeasurably high friction)} was reported for various fluids on mica, in spite of the \MLR{atomic} smoothness of the interface \cite{chan1985drainage,maali_measurement_2008}.
%we chose the atomically smooth and insulating (001) plane of muscovite mica. 
%It is interesting
Here we specifically designed a tuning-fork based atomic force microscope (AFM) system allowing us to measure the slip length of glycerol, with nanometric resolution, as a function of temperature. 

We will then put our experimental results into perspective with predictions of molecular dynamics (MD) simulations of model supercooled liquids. 
%{The mechanisms of surface slippage and friction of supercooled liquids has been recently investigated by molecular dynamics (MD) simulations of model systems} and 
These simulations suggest an increased slippage in the supercooled regime of water, ethanol \cite{herrero_fast_2020} or binary Lennard-Jones liquids \cite{lafon_giant_2022}, typically by a factor up to 5 depending on temperature. The authors furthermore reported an Arrhenius-like \textit{decrease} of friction as temperature increases, {in agreement with the picture that an increased thermal agitation accelerates liquid dynamics and thus reduces friction.} 
%Let us 
Interestingly however, these simulations use frozen walls and neither consider the internal degrees of freedom of the solids, nor their couplings to the liquid. These numerical results thus act as benchmark predictions for surface friction without the solid internal dynamics. 

\begin{figure*}
\centering
\includegraphics[width=\linewidth]{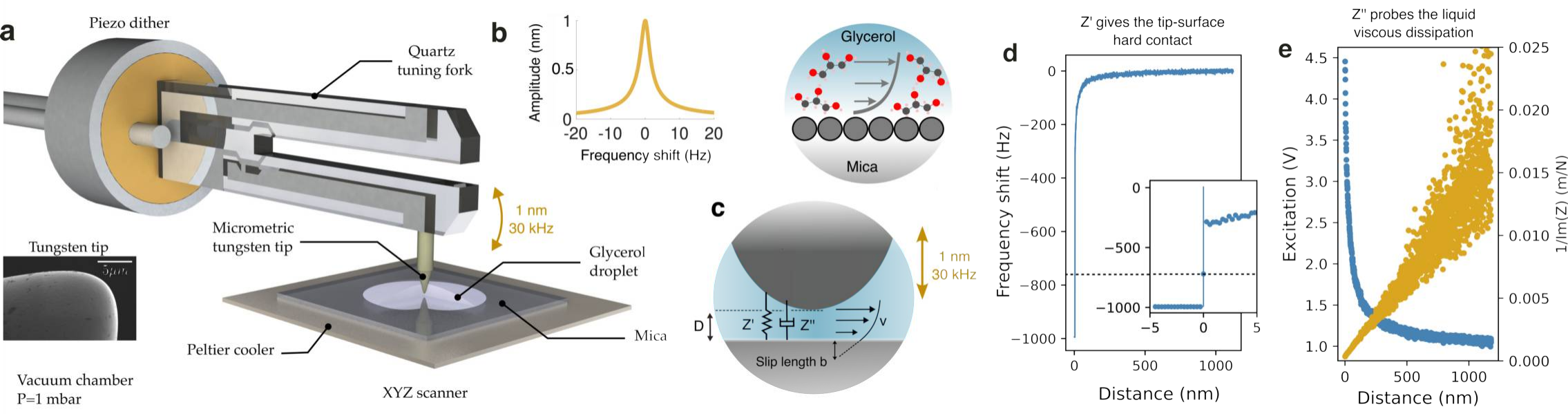} %{Figure1.png} 
\caption{\label{figure1}\textbf{AFM slippage measurement:} \textbf{a} schematics of the tuning fork AFM system, with a typical resonance measurements shown on panel \textbf{b}; \textbf{c}: sketch of the drainage flow between the micrometric tungsten tip and the mica surface. 
%\LB{ \sout{ Glycerol viscosity versus temperature is shown on \textbf{c} : the thick blue curve is obtained experimentally with a cone-plane rheometer. The black line denotes glycerol's viscosity with a 0.5 \% water contamination \cite{cheng_formula_2008}. The thin blue lines are the real and imaginary part	s of glycerol viscosity at 30 kHz.}} 
\textbf{d} Elastic response $Z^\prime$ of the confined liquid. The inset shows a zoom on the mechanical contact where the phase regulation is suddenly lost and the frequency shift saturates at -1 kHz. This discontinuity allows us to determine the position of the mechanical contact ($D=0$) with nanometric resolution. \textbf{e} Typical approach curve at room temperature (blue) showing the dissipation increasing as more energy is lost to the confined viscous flow. In orange, we show the inverse dissipative impedance $1/\Zim$ versus sphere-plane distance $D$.
}
\end{figure*}

\subsection*{Experimental setup and results}
Our measurements of liquid-solid friction are based on an axi-symetric flow between an AFM tip and a mica surface as sketched on Figure \ref{figure1}\textbf{a-b}. A tungsten tip of micrometric diameter is glued on the quartz tuning fork and immersed in a glycerol droplet placed on a freshly cleaved mica surface. The prong is excited at its resonant frequency of roughly 30 kHz with a piezo dither, leading to a vertical oscillatory motion of the tip. The tuning fork's oscillation signal is measured with a trans-impedance preamplifier (FEMTO DLPCA-200). This signal is then directed to a phase-locked-loop (Nanonis Mimea) \ML{which keeps the phase to zero and the system at resonance} (\textit{cf.} inset of Figure \ref{figure1}\textbf{a}) and monitor the excitation voltage to maintain a constant nanometric amplitude ($a = 3$ nm).

\ML{The drainage flow between the tip and mica induces forces which cause a slight change of the resonance curve: elastic forces that are in phase with displacement, shift the fork's resonance frequency by $\delta f$. On the other hand, viscous friction damps the oscillation.} This damping is compensated by an increase of the piezo excitation ($E$) to maintain the constant oscillation amplitude $a$. With this measurement scheme, we obtain the complex mechanical impedance $Z$ of the confined liquid \cite{comtet2018rheology} (here at \ML{30 kHz excitation ; \textit{cf.} Figure \ref{figure1}\textbf{b}):

\begin{equation}\label{impedance_exp}
	Z^{\prime}+iZ^{\prime \prime}= 2K_{\textrm{eff}}\frac{\delta f}{f_0}+i\frac{\Keff}{Q_0}\Bigl(\frac{E}{E_0}-1\Bigr)
\end{equation}

where $K_{\textrm{eff}}$ is the effective oscillator's stiffness, $f_0$, $Q_0$ and $E_0$ are respectively the free frequency, free quality factor and free excitation voltage. We stress that the very high stiffness ($\approx$40 kN) of this system makes it very robust with respect to strong forces that typically cause the tip to snap-in to the surface in AFM experiments. Moreover, ensuring that only the very end of the tip is in glycerol yields an excellent quality factor and thus an exquisite resolution on $Z$.}

%We stress that the very high stiffness ($\approx$40 kN) of this system makes it very robust against strong force gradients typically occurring close to mechanical contact. 
%Moreover, placing only the very end of the tip in glycerol yields an excellent quality factor and thus an exquisite resolution of the mechanical impedance.

\LB{The mechanical contact} between the AFM probe and the mica surface is detected as a divergence of the frequency shift $\delta f$ (contact stiffness) shown on Figure \ref{figure1}\textbf{d}. This sharp and reproducible divergence \ML{reliably provides the origin of the x axis ($D=0$) with nanometric resolution}. E and $1/\Zim$ are plotted on Figure \ref{figure1}\textbf{e}, as functions of the sphere-plane distance $D$ \ML{for an approach curve measured at room temperature.} %Here, the Reynolds prediction that $1/Im(Z)$ is linear in D is expected to hold for purely viscous liquids in the absence of slippage.

The mica sample is placed on a Peltier element allowing to control its temperature. To avoid any condensation of water in the glycerol at the lowest temperatures, we place the whole system in a vacuum chamber under a pressure of 1 mbar, \ML{above the vapour pressure of glycerol but below that of water}. The chamber can also be filled with dry nitrogen and the pumps turned off during measurements to decrease the mechanical noise level. All curves presented here have been measured with the same tip (13 $\mu$m in diameter, \ML{see Figure \ref{figure1}\textbf{a}), immersed in the droplet for several weeks. Below its melting temperature $T_M$ = 18$^{\circ}$C, glycerol is in the supercooled regime and its viscosity increases continuously over a dozen of orders of magnitude when approaching the glass transition at $T_g \sim$ -87$^{\circ}$C \cite{schroter2000viscosity,davies1973viscoelastic}. We used a cone-plane rheometer to assess the purity of our glycerol and found its viscosity is in excellent agreement with the literature \cite{cheng_formula_2008}, showing that our water contamination level is inferior to 0.5$\%$ in mass (\textit{cf.} Section I.A in SI). Measurements of the complex shear modulus at 30 kHz show that glycerol's visco-elasticity is negligible for $T>0^\circ$C (\textit{cf.} Section I.A in SI). In other words, hydrodynamic and molecular timescales are well separated.}% $\vert\mathfrak{Im}(\eta)/\mathfrak{Re}(\eta)\vert < 30$.}

\begin{figure*}
	\centering
	\includegraphics[width=\linewidth]{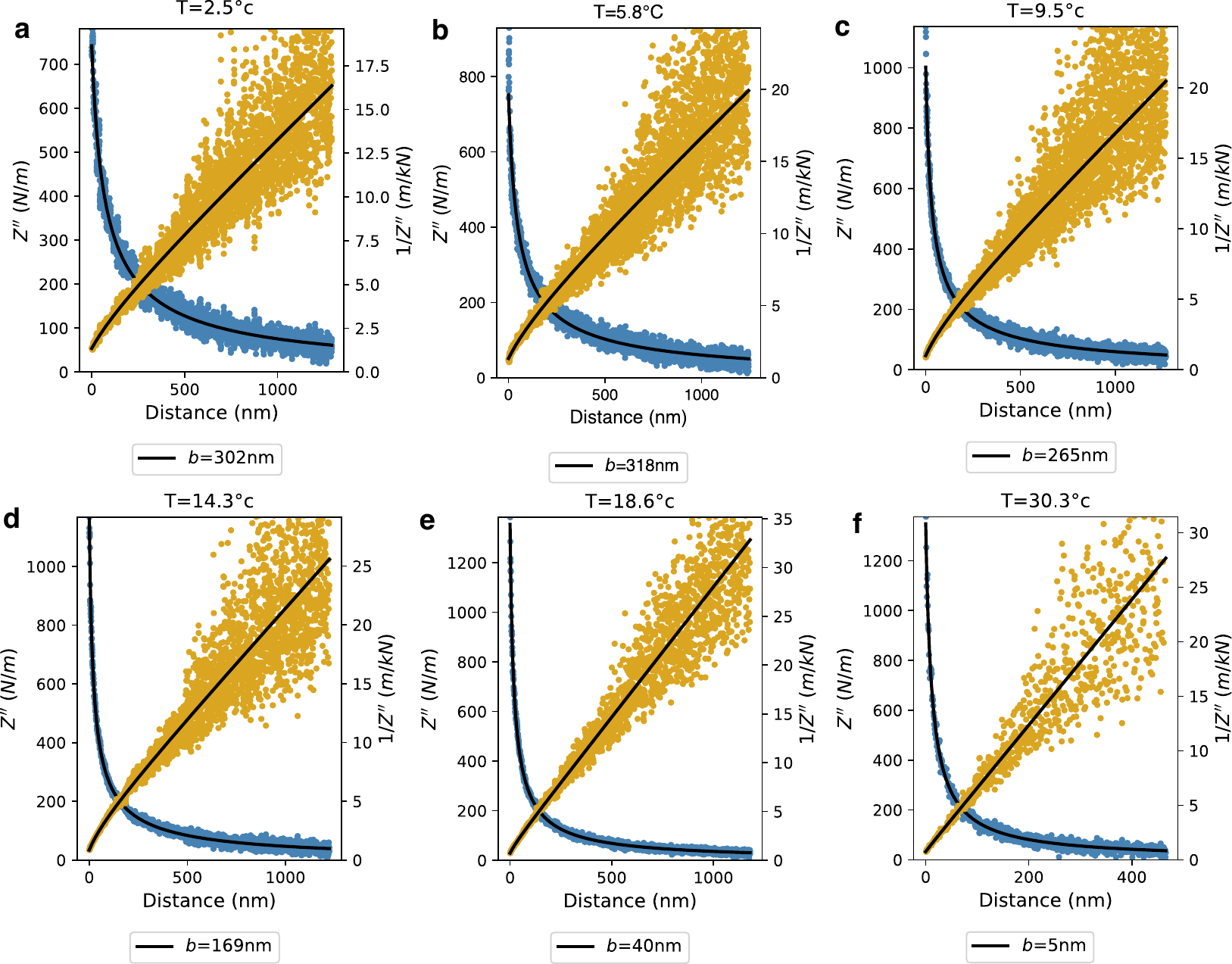} 
	\caption{\label{fig2}\textbf{Temperature dependent friction:} \textbf{a-d} Inverse dissipative impedance $1/\Zim$ as a function of sphere-plane distance $D$. The solid lines are fits using Eq.\ref{vinogradova} for $T\in \{2.45,5.8,9.5,14.3,18.6,30.3\}$$^{\circ}$C (see SI Section II.A for details on the fitting process). }
\end{figure*}

\ML{In the presence of a slip length b at the surface and in the sphere-plane geometry, the dissipative mechanical impedance, $\Zim$ is reduced with respect to the no-slip case and takes the Reynolds form at long distance \cite{cottin-bizonne_nanohydrodynamics_2008}}

\begin{equation}\label{reynolds}
    \Zim_{Reynolds} = \mathfrak{Im}(Z) = \frac{6  \pi R^2 \eta\, \omega}{\LB{D+b}}, %\Bigl(1-\frac{b}{D}\Bigr)
\end{equation}
\ML{where $1/\Zim$ is a straight line. Note that we assume a no-slip boundary condition for the relatively rough tungsten tip. At short distances, $D \leq b$, $1/\Zim$ deviates from linearity as slippage induces strong deviations from the Reynolds equation and a more general formula is needed.}

\begin{equation}\label{vinogradova}
    \Zim=\frac{6\pi R^2 \eta \omega}{D}f^*\Bigl(\ \frac{b}{ D}\Bigr) %\frac{\tilde{\eta}}{\tilde{\lambda} D}\Bigr)
\end{equation}
 \LB{The expression for $f^*$ was previously derived in the literature \cite{vinogradova_drainage_1995,cottin-bizonne_nanohydrodynamics_2008}}  \ML{(\textit{cf.} Section II.A in SI) and is used to fit our approach curves in a systematic way.}

This expression \LB{discards} hydrodynamic inertia effects since the viscous penetration length ($\sqrt{\eta/\rho\omega}>100 \mu m$) is much larger than the probe's radius $R\sim6.5\mu$m. It was recently shown that inertial effects have negligible influence on $Z^{\prime \prime}$ in this regime \cite{zhang2023unsteady} allowing us to neglect them in the following \ML{(\textit{cf.} Section II.E in SI)}. Nevertheless, and still following \cite{zhang2023unsteady}, it is interesting to note that they are \LBB{present} for the elastic part of the impedance $Z^{\prime}$ and yield the negative frequency shift reported in Figure \ref{figure1}\textbf{d}. In Eq.\ref{vinogradova}, we also neglect the elastic deformations of the solid surface leading to elasto-hydrodynamic corrections, \ML{as mica does not deform under the flow in present conditions {(\textit{cf.} Section II.C in SI)}.}

\LB{Let us first examine the high temperature regime (here, close to room temperature). As shown} on Figure \ref{fig2}\textbf{e}, the experimental results are in very good agreement with the usual Reynolds prediction in Eq.(\ref{reynolds}): \ML{the $1/\Zim$ versus $D$ line crosses the x-axis close to $D=0$, showing that the slip length $b$ is vanishingly small. }

\LB{Now, as the temperature decreases from $30^\circ$C down to $1^\circ$C, one observes that the general shape of the curves for $1/Z^{\prime \prime}$ versus $D$ slightly change and the standard Reynolds expression in Eq.(\ref{reynolds}) is not sufficient to describe the dissipation for lower temperatures. However, the generalized prediction in Eq.(\ref{vinogradova}) is shown to be in good agreement with the experimental data, provided a non-vanishing slip length $b$ is introduced; see Figure \ref{fig2}\textbf{a}-\textbf{e}.} \ML{By fitting the dissipation curves (blue points in Figure \ref{fig2}) with Eq.\ref{vinogradova} (see details in SI Section II.A), one can accordingly extract the slip length $b$ over a wide range of temperatures. The model obtained from the fit is then plotted as a black solid line on both the $\Zim$ (blue) and $1/\Zim$ (orange) points.}
%\ML{We can now} compare the inverse dissipative impedance $1/\Zim$ \ML{curve} with \ML{the corresponding fit with} Eq.\ref{visco_elastic_general} for $T\in \{2.45,9.5,14.3,18.6,30.3\}$$^{\circ}$C (black solid line fits).
\ML{Clearly, Eq.\ref{vinogradova}} reproduces very precisely the \LB{distance-dependence} of the inverse approach curves $1/\Zim$. The error on the slip length from a single fit is \ML{of the order of 3 nm} for all temperatures.

\begin{figure*}
	\centering
	\includegraphics[width=\linewidth]{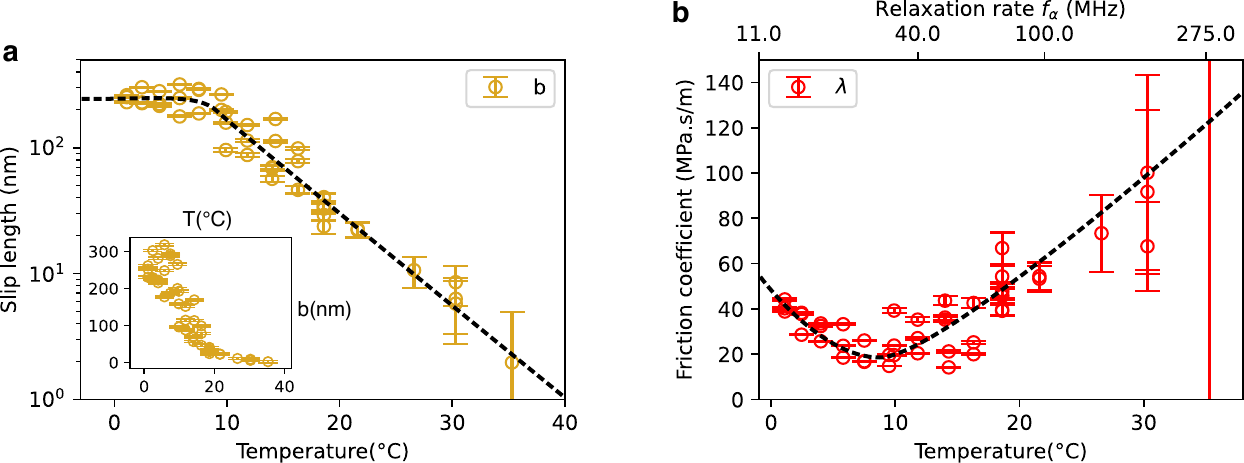} 
	\caption{\label{fig3}\textbf{Interfacial friction and slip length \textit{versus} temperature:} All measurements are done in a single experimental run for various temperatures, with the same tip and on the same mica surface. \textbf{a} Dynamical slip length $b$ versus temperature with linear (\textit{inset}) and log scales (main panel).  \textbf{b} Resulting interfacial friction coefficient, defined as $\lambda(T)=\eta(T)/b(T)$,  versus temperature ${\lambda} (T)$. We also give the liquid's $\alpha$-relaxation frequency, $f_\alpha(T)$ on the top x-axis.}
\end{figure*}

\subsection*{Discussion and possible dissipation mechanisms}

\LBB{Gathering results, we} plot on Figure \ref{fig3}\textbf{a} the extracted slip length $b$ as a function of the glycerol temperature between 0$^{\circ}$C and 35$^{\circ}$C. We immediately notice a \LBB{two orders of magnitude} increase of slippage upon cooling with $b$ going from 0-2 nm at 35$^\circ$C to $\sim$200 nm at 1$^\circ$C.
%We find a dramatic increase of slippage upon cooling. While $b$ is nanometric at high temperature - of the order of our instrumental resolution - it reaches hundreds of nanometers below 0$^{\circ}$C. 
We emphasize that such a behavior is not expected and contrasts with theoretical expectations and MD simulations \cite{bocquet_nanofluidics_2010,lafon_giant_2022}. Indeed, the slip length $b=\eta/\lambda$ of simple liquids is expected to be independent of viscosity \cite{bocquet_nanofluidics_2010}, in \MLR{accordance with the few experiments we are aware of \cite{cottin2008nanohydrodynamics}. This is due to the similar linear dependence on the liquid's molecular timescale for surface friction $\lambda$ on the one hand and viscosity $\eta$ on the other.} Accordingly $b$ would be expected here to depend only weakly on temperature. Such a behavior is \MLR{indeed} observed in MD simulations far above the glass transition \cite{herrero_fast_2020,lafon_giant_2022}, where an increase at most by a factor 5 is measured. In our experiments we do in fact observe a saturation of the slip length as a function of temperature \textit{below} $10^\circ$C. 
However, the huge decrease of slip length measured for $T>10^\circ$C \MLR{defies} expectations.

Digging further, we plot the experimental friction coefficient $\lambda(T)$ \textit{versus} temperature on Figure \ref{fig3}\textbf{b} \MLR{-- see SI Section I for viscosity measurements.} The friction coefficient is shown to exhibit a non-monotonous dependence on temperature, with a minimum \MLR{around $10^\circ$C and an increase} of $\lambda(T)$ for $T>10^\circ$C. This behavior is at odd with \MLR{the usual Arrhenius picture} for $\lambda(T)$, which predicts an overall decrease of $\lambda(T)$ with T. The latter behavior is indeed  observed in MD simulations of binary Lennard-Jones liquids, water or methanol in the supercooled regime \cite{herrero_fast_2020,lafon_giant_2022}.

One may also notice that the strong increase of friction with temperature \LBB{above $10^\circ$C,} \MLR{may seem disproportionate to the relatively small variation in thermal energy} ($\sim$ 10$\%$). However, even \MLR{in} this moderate temperature range, glycerol dynamics are varied by a factor {of 30, in agreement with the drastic increase of viscosity from 0.5 to 12 Pa.s.} This is best evidenced in the strong \MLR{redshift} of the characteristic $\alpha$-peak of glycerol $f_\alpha(T)$ \MLR{upon cooling} ; see Supplementary \MLR{Section I}. Accordingly, in order to highlight the relationship between friction and glycerol dynamics, %uncover the mechanism behind this friction reduction, 
we \MLR{convert temperature into the corresponding frequency $f_\alpha(T)$ on the top axis of Figures \ref{fig3}\textbf{b}}. This plot highlights two regimes : for $f_{\alpha} < 34$ MHz, friction is measured to \MLR{be} \LB{decreasing with the molecular frequency $f_\alpha$, down to $\lambda_0 \sim 20$ MPa.s/m}. For $f_{\alpha} > 34$ MHz, however, the friction coefficient increases with the molecular frequency $f_{\alpha}$. 

Such a behavior has neither been reported experimentally up to now,  nor explained theoretically. It strongly differs from the predictions of MD simulations, with frozen confining walls \cite{herrero_fast_2020,lafon_giant_2022}. One can explore various possible mechanisms to explain it.
%\LBB{{This suggests that other mechanisms are at play in the experiments.}}

\subsubsection*{Strain rate dependency and non-linearities}
%\paragraph*{Strain rate dependency --}
For liquid slippage, non-linearities may take the form of a strain-rate dependency where the strain rate $\dot{\gamma} \sim \partial_z v_x$ is a typical frequency scale of the flow.  Such non-linearities are believed to emerge when microscopic timescales and strain rate are of the same order of magnitude. In our experiments, the strain rate is not constant during the approach but we can estimate it roughly as  $\dot{\gamma} \sim 2\pi f  a/D \sim 10$ kHz. We know from molecular dynamic simulations that both viscosity and friction coefficient are constant for $\dot{\gamma} \tau < 0.1$ where $\tau$ is a typical liquid's relaxation rate \cite{priezjev2004molecular}. In the worst case of $T=0^{\circ}$C, this criterion yields a critical shear rate of roughly 100 kHz, more than one order of magnitude higher than what is reached experimentally. These considerations lead us to ignore the nonlinear effects caused by strain rate dependency of the friction coefficient.

\subsubsection*{Interfacial water}
%\paragraph*{Interfacial water --}
\LB{One may} also consider the possibility that water or gas could condense \LB{preferentially} at the hydrophilic mica surface. Considering the large affinity of glycerol with water, we were extremely careful to avoid water contamination: the 99.9$\%$ pure glycerol was deposited on freshly cleaved mica and immediatly pumped down to 1 mbar \LB{in the experimental cell} for several days. Still, one can never completely rule out  the possibility that water could condense at the mica surface. In such a case, this low viscosity lubrication layer \LB{at the mica interface would} yield an apparent slippage \LB{proportionnal to the viscosity ratio, as $ b_{app} \sim (\eta_{glycerol}/\eta_{water}) \times \delta$, with $\delta$ the water interfacial thickness} \cite{andrienko2003boundary}. %Assuming no-slip conditions for a worst case estimate, 
\LB{The experimental results shown in Figure \ref{fig3} would imply a (temperature dependent) thickness of the water layer in the range of} $ \delta  \sim 10^{-2}$ nm, which is hardly realistic. We thus conclude that water segregation at the interface \LB{can be discarded as an underlying mechanism}. %One could argue that a partial coverage of the surface yields a small effective thickness $\delta$. In such a scenario, we expect a spatial heterogeneity of slippage which was not observed. To account the experimental data, the coverage (effective $\delta$) should increase with decreasing temperature. Such a prewetting upon cooling should also be reversible to account for the absence of hysteresis in the measurements.

\subsubsection*{Solid internal dynamics and fluctuation-induced dissipation}
%\paragraph*{Solid internal dynamics --}
Putting our experimental results in perspective with the MD simulations of supercooled liquid friction -- which report an Arrhenian behavior \cite{herrero_fast_2020,lafon_giant_2022}  and hence do no account for the non-monotonous dependence of friction  with temperature --, one key ingredient which is missingin the MD modelling is the internal dynamics of the solid. In contrast, recent MD calculation of liquid-solid friction including internal degrees of freedom of the solid showed that vibrational coupling may affect the friction coefficient \cite{noh2022effect}. 
 It is therefore tempting to investigate the role of wall internal fluctuations on surface friction. Fundamentally, liquid friction can be calculated from the fluctuation-dissipation theorem using the Green-Kubo formula :
\begin{equation}\label{green_kubo}
	\lambda = \frac{1}{\mathcal{A} k_B T}\int_0^\infty \langle F_x(t)F_x(0)\rangle
\end{equation}
\LBB{with $\cal{A}$ the lateral area and $F_x$ is the fluid-solid interaction force.}
\LB{Let us assume that the fluid-solid interaction is described by an interaction potential $V_{FS}$. The surface force can be accordingly rewritten in terms of the (fluctuating) density distribution in the liquid and the solid as $F(t) = \int dr_s dr_l \nabla V_{FS} (r_s-r_l) n_s(r_e) n_l(r_w)$.}
%Starting from the Green-Kubo Eq.\ref{green_kubo}, we assume that forces derive from the Lennard-Jones potential and density distributions in the liquid and solid : $F(t) = \int dr_s dr_l \nabla V_{LJ} (r_s-r_l) n_s(r_e) n_l(r_w)$. 
Considering that fluctuations in the solid and in the liquid are uncorrelated at first order, we can  rewrite the Green-Kubo equation \LB{in terms of} the interfacial (2D) structure factors $S(r,r',t,t') = \langle n(r,t)n(r',t')\rangle$ of the two media. 
\LB{Going to Fourier space allows us to rewrite the friction coefficient in Eq.(\ref{green_kubo}) }
%We Fourier transform in space and time and rewrite the friction coefficient 
as a function of the angle-averaged dynamical structure factors of the solid $S_s(q,\omega)$ and liquid $S_l(q,\omega)$: 
\begin{equation}
\label{friction_structure} \lambda = \frac{1}{8 \pi^2 k_BT}\int_0^{\infty} dq V_{FS}(q)^2 q^3 \int_0^{\infty} d\omega S_s(q,\omega)S_l(q,\omega) 
\end{equation}
%We underline that this friction estimate from the Green-Kubo relation is based on a quasistatic assumption and assumes a vanishingly small flow velocity. 
We now estimate $\lambda$ by identifying the key contributions to the solid and liquid fluctuation spectra. The full derivation is reported in the \MLR{SI Section III} and we discuss here the main steps.
%As a side note, we quote that this theoretical description does not involve by simplicity the visco-elasticity of the liquid, and the resulting friction coefficient is a real quantity. 
%\LB{[reprendre]This yields a real friction coefficient \textit{by construction} in contrast with the experimentally obtained complex value reflecting the interfacial shear rheology at 30 kHz.  All comparison between this theoretical derivation and experimental results from Figure \ref{fig3} shall therefore be taken with a grain of salt.}

\paragraph*{Liquid's structure factor --}
\LB{Spectroscopy measurements of bulk glycerol -- in particular dissipative mechanical and electrical impedance ($G''(\omega)$ and $\epsilon''(\omega)$) \cite{jensen2018slow,davies1973viscoelastic,lunkenheimer_high-frequency_1996} -- show that glycerol exhibit a non-dispersive Debye-like $\alpha$-peak at a temperature dependent frequency $f_{\alpha} \sim e^{-E_a/k_B(T-T_g)}$, with $T_g$ the glass temperature transition. }
%To further infer the liquid's dynamical structure factor, we use measurements of the dissipative mechanical and electrical impedance ($G''(\omega)$ and $\epsilon''(\omega)$) of bulk glycerol, respectively measured with mechanical \cite{jensen2018slow} or dielectric spectroscopy \cite{lunkenheimer_high-frequency_1996}. 
%In analogy with the structure factor of water, we model the glycerol's $\alpha$ relaxation as a non dispersive Debye peak centered at the temperature dependent frequency $\omega_{\alpha} \sim e^{-E_a/k_B(T-T_g)}$ of the dielectric loss peak \cite{lunkenheimer_high-frequency_1996}. 
For the sake of simplicity and considering the dispersionless nature of the peak, we define a effective liquid's structure factor \LBB{whose frequency dependence is Debye-like:}%(here, $q$-averaged): 
\begin{equation}\label{liquid_structure} S_l(\omega) = \frac{\omega_{\alpha}(T)}{\omega_{\alpha}(T)^2+\omega^2}
\end{equation}
\LBB{We refer to the Supplementary Information for further details ($q$ dependence, etc.).}
% In the computation, we used the full wavevector dependency of $S_l$ (see Appendix) which assumed to be frequency independent and shows a cut-off for wavevectors larger than the inverse molecular size of glycerol.
%where the static part can be written as \begin{equation}\lambda^{static} =  \frac{S(q_{\parallel})\rho_c}{2\Dqparr k_B T} \end{equation} with $\Dqparr$ the collective interfacial diffusion coefficient, $\rho_c$  the fluid's density and $S(q_{\parallel})$ its in-plane static structure factor at the surface's typical wavevector.

\paragraph*{Solid's structure factor --}
The solid's structure factor \LB{can be separated} into a non-fluctuating (static) part that accounts for static \LB{corrugation} and a dynamical part that accounts for internal fluctuations and excitations, $S_s = S_s^{\rm static} + S_s^{\rm dyn}$. This yields two complementary contributions to friction: $\lambda = \lambda^{static} + \lambda^{dyn}$. The solid's static structure factor ($\omega = 0$) that accounts for roughness is strongly peaked at the reciprocal lattice typical wavevector \LB{$q_s^{\textrm{max}}\approx \frac{2\pi}{\sigma_s}$with $\sigma_s$ a typical mica molecular length scale, so that } 
\begin{equation}
S_s^{\rm stat}(q,\omega)\approx u_s\rho_s q_s^{\textrm{max}}\delta\left(q-q_s^{\textrm{max}}\right)\delta(\omega)
\end{equation}
where $\rho_s\approx 1/\sigma_s^2$ is the atomic density on the interacting layer and $u_s$ is the amplitude of the roughness. 
\LB{The dynamical part of the  fluctuations spectrum, $S_s^{\rm dyn}$, accounts for the internal excitations. As an insulator, mica does not exhibit low-energy electronic excitations, but }
%As an insulator, mica is unable to sustain low-energy electronic excitations. Nonetheless several 
phonon modes of mica were measured by Brillouin scattering \cite{mcneil1993elastic}. \LBB{Keeping the description as simple as possible,} we model the dynamical structure factor of mica using one single acoustic phonon branch \LB{as} 
\begin{equation}
S_{\rm ph}(q,\omega)%=\pi \frac{T\rho_s q^2}{m\omega^2}\delta (\omega\pm q\cdot c)
=\pi \frac{T\rho_s}{mc^2}\delta (\omega\pm q\cdot c)
\end{equation}
Considering the aforementioned Brillouin scattering experiments, we estimate the sound velocity to be of the order of $c\approx 10^3$ m/s \cite{mcneil1993elastic} while mica's lattice parameter is in the nanometer range \cite{pintea2016solid}.% and write the corresponding dynamical structure factor: \begin{equation} S_{s}^{dyn}(q,\omega) \simeq \delta(\omega \pm qc) \frac{\pi q k_B T}{\hbar c}\end{equation} where c is the phonon's group velocity.

%Turning to the dynamical contribution, we estimate the dynamical structure factors $S_s$ and $S_l$ from dielectric response and the fluctuation-dissipation relation.  At zero-wavevector, the dielectric response of glycerol $\epsilon''(\omega)$ shows two main contributions : a slow $\alpha$ relaxation at $\omega_{\alpha}$ which strongly shifts with temperature and a high frequency, temperature insensitive boson peak \cite{lunkenheimer_glassy_2000} We assume an exponential relaxation on scale $q_0$ and write a reasonable ansatz for $\epsilon''(q,\omega)$. The $\alpha$ relaxation controls the liquid's viscosity and can be approximated by a Debye peak with $1/\omega$ tails:
%\begin{equation}\epsilon''_{\alpha}(q,\omega) = g_0 e^{-q/q_0}\frac{\omega_{\alpha}}{\omega_{\alpha}-i \omega}\end{equation} where the peak frequency $\omega_{\alpha} \sim e^{-E_a/k_B(T-T_g)}$ \cite{lunkenheimer_high-frequency_1996} and where $g_0$ is independent of temperature. %The minimal assumption of a single non-dispersive mode at frequency $\omega_p$ in the solid leads to: \begin{equation} \lambda^{dyn}_{plasmon} \sim  \Gamma k_B T \frac{q_{max}^3\omega_{\alpha}}{\omega_p^2}\end{equation} where $q_{max}$ is fixed either by glycerol's or the solid's structure factor.

%The minimal assumption of a single acoustic phonon for the mica ($\omega_p = cq$) yields: \begin{equation} \lambda^{dyn}_{phonon} \sim  k_B T \frac{q_{max}^2\omega_{\alpha}}{8 \pi c^2}\end{equation} and thus accounts for the experimental scaling $\lambda^{dyn}\propto \omega_{\alpha}$. 

%\LB{ICICICI}

\begin{figure}%\centering
	\includegraphics[width=12cm]{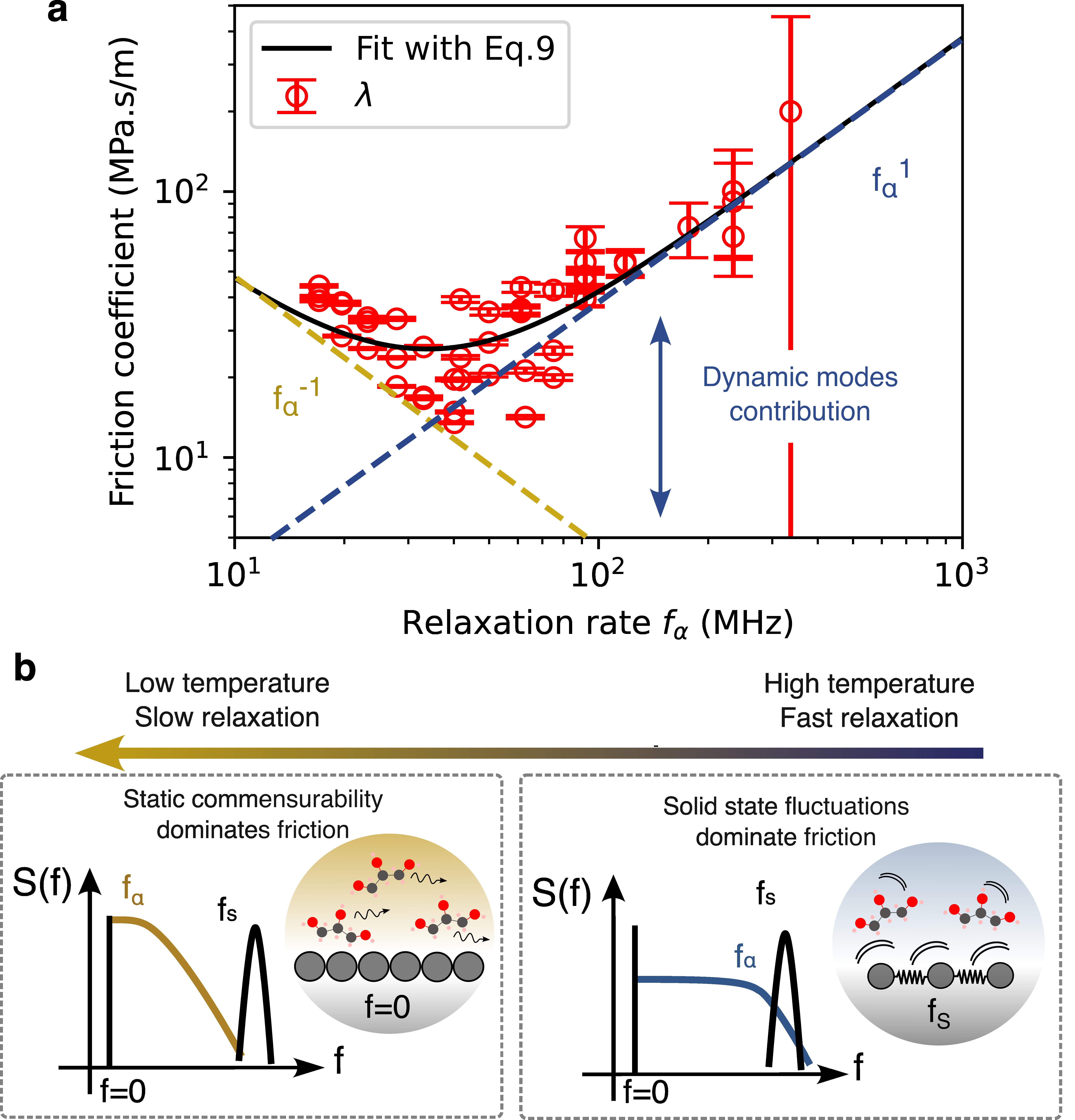} 
	\caption{\label{figure4} \LB{{\bf a} Parametric plot of the experimental values of the solid-liquid friction coefficient, $\lambda(T)$, versus the (temperature-dependent) frequency of the $\alpha$-peak, $f_\alpha(T)$. The solid line is the theoretical prediction in Eq.(\ref{friction}).
	%$\lambda(\omega_\alpha)=\lambda_0\times \left({\omega_c\over\omega_\alpha}+{\omega_\alpha\over\omega_c}\right)$. 
	The fit to the experimental data yields $\lambda_0=25.6\cdot 10^6 {\rm N.s/m}^3$ and $f_c=33.9$MHz.   %Comparison of the theoretical prediction with experimental data for the modulus of the friction coefficient $\vert \lambda \vert$.
	{\bf b} Sketch of the underlying physical mechanisms, \LBB{highlighting an increased dynamical dissipation when the fluid spectrum overlaps the solid internal spectrum. The solid peak at zero frequency accounts for the static roughness of the solid, while the one at $f_S$ (for a given $q$) is associated with the dynamical solid fluctuations.}
	}}
\end{figure}

%\paragraph*{Two contributions to friction--}
\paragraph*{Frequency scalings and discussion}
\LB{Using these expressions for the spectra in Eq.(\ref{friction_structure}), one can calculate the 'static' (corrugation induced) and 'dynamic' (solid fluctuation induced) contributions to the friction coefficient. Note that for the sake of simplicity and in order to proceed with calculations, we used a simple Lennard-Jones interaction potential for the fluid-solid interaction to obtain analytical estimates, although the main qualitative results do not
depend on this hypothesis. We leave the detailed derivation in the Supplementary Information. \LBB{A key prediction emerging from this calculation is that the friction
coefficient is found to obey a simple functional dependence on the liquid molecular dynamics -- characterized by its $\alpha$-peak frequency $f_\alpha$ -- in the general form}
\begin{equation}
\lambda (f_\alpha)={\lambda_0\over 2} \left( {f_c\over f_{\alpha}}+{f_\alpha \over f_{c}}\right)
\label{friction}
\end{equation}
with the two terms originating in the 'static' and 'dynamic' contributions to the interfacial friction, respectively. The expressions for $\lambda_0$ and $f_c$ as a function of the molecular parameters are provided in Supplementary Information.
%with scalings as
%\begin{equation}
%\lambda^{\rm stat} \propto {1\over \omega_{\alpha}}
%\end{equation}
%and
%\begin{equation}
%\lambda^{\rm dyn} \propto {\omega_{\alpha}}
%\end{equation}

As shown in Figure \ref{figure4}, the prediction in Eq.(\ref{friction}) is in good agreement with the experimental data for the friction coefficient and its functional dependence on the molecular dynamics via $f_\alpha(T)$. The fit yields the values $f_c\approx 33.9$MHz as the threshold frequency and $\lambda_0\simeq 25.6\cdot 10^6$N.m/s. The scaling for the first (static) contribution, $\lambda^{\rm stat}\propto 1/f_{\alpha} $, is expected} for the friction on solids with a static roughness since the friction coefficient in this regime is merely proportional to the fluid relaxation time-scale, hence as $1/f_{\alpha}(T)$,  in agreement with previous calculations \cite{bocquet_nanofluidics_2010}. This is consistent with the measured slip length being independent of the temperature for $T<10^\circ$C, see Figure \ref{fig3}\textbf{a}. \MLR{In this regime, as sketched on Figure \ref{figure4}\textbf{b}, the structure factors overlap is dominated by the solid's static corrugation at zero frequency, the influence of mica's phonons is thus negligible.} At fast relaxation rates however, the scaling of the
dynamical contribution to friction $\lambda^{\rm dyn}(T)\propto {f_{\alpha}(T)}$ \MLR{contradicts the Arrhenius picture and stems from the high overlap of liquid's structure factor $S_l$ with solid excitations $S_s^{\rm dyn}$. The sketch of Figure \ref{figure4}\textbf{b} highlights this crossover: above $f_c \sim 34$ MHz, the liquid's coupling with mica's phonons %(at $f_{\rm s}$) 
dominates that with the static corrugation at 0 Hz. }%is counter-intuitive, and stems from the increase of excitations' population with temperature, hence its associated dissipation.

%Using Eq.\ref{friction_structure}, we now estimate both the static and dynamic friction coefficients and focus on their relative amplitude as a function of $\omega_{\alpha}$. The static wall contribution sits in the $\omega \ll \omega_{\alpha}$ limit where $S_l(\omega) \sim 1/\omega_{\alpha}$ (\textit{cf.} Eq.\ref{liquid_structure}) thus leading to $\lambda^{stat} \propto 1/\omega_{\alpha}$. 
%\LB{The prediction in Eq.\ref{friction_structure} suggests that the} solid wall's fluctuations will contribute to the interfacial friction \LB{the liquid and solid fluctuation spectra overlap for $\omega \neq 0$.} 
%% match the liquid's structure factor, \textit{ie.} if there is overlap between $S_s$ and $S_l$ for $\omega \neq 0$. 
%
%The dynamical contribution from the acoustic phonon branch is in the opposite limit $\omega \gg \omega_{\alpha}$ in which case $\lambda^{stat} \sim S_l(\omega) \sim \omega_{\alpha}$. Consequently, we obtain two opposite scalings for the friction coefficient, \begin{equation} \lambda^{stat} \propto 1/\omega_{\alpha} \textrm{   and    } \lambda^{dyn}\propto \omega_{\alpha}\end{equation}  for respectively the low and high frequency regimes. 
%
%Interestingly, the friction drop with increasing $\omega_{\alpha}$ expected for $\lambda^{stat}$ is observed  experimentally in the imaginary part of friction only $Im(\lambda)$ (\textit{cf.} Figure \ref{fig3}\textbf{d}).  For its part, the linear scaling of $\lambda(\omega_{\alpha})$ is clearly apparent for $Re(\lambda)$ on Figure \ref{fig3}\textbf{c}. 
Altogether, the low and high temperature regimes highlighted in the experimental data can therefore be interpreted as respectively the static (corrugation-induced) and dynamic (due to internal solid fluctuations) contributions to hydrodynamic friction. The cross-over between the two regimes occurs at a characteristic frequency $f_c$, which separates the static and dynamic frictional regimes.

The description therefore captures the main features of the experiments. However, we mention that estimating more quantitatively the measured values for $\lambda_0$ and $f_c$ (see above) on the basis on their predicted molecular expressions -- as reported in the Supplementary Informations -- is difficult because the latter depend very sensitively on various molecular details. It would also require to go beyong the simple theoretical description of the liquid and solid spectra, as well as the simplistic (Lennard-Jones type) molecular interactions that we use here. More fundamentally, the dynamics of supercooled liquids close to surfaces was shown to be strongly affected by surfaces, with drastic change of relaxation time-scales depending on the wall roughness \cite{scheidler2002cooperative,kob2012non}. This interfacial effect may affect the quantitative estimate of $f_c$ by orders of magnitudes, but it remains difficult to predict quantitatively.
Hence we leave a full quantitative analysis for future work using proper molecular simulations and keep here our discussion to a qualitative - scaling - analysis which reproduces the main features of the experimental results. 

As a last comment, the interfacial friction coefficient $\lambda_0$ on mica is measured to be very large (in the tens of MN.m/s), and only a viscous fluid -- like glycerol here -- could exhibit slippage on a surface with such high friction. This is consistent with the  no slip boundary condition previously measured for water and OMCTS on mica \cite{chan1985drainage,maali_measurement_2008}.

\subsection*{Conclusions}

In this work, we introduced a tuning-fork based AFM optimized for liquid-solid friction measurement in a wide range of viscosity and temperature, both in vacuum and in controlled atmosphere. We use this new system to explore the temperature-dependent slippage of supercooled glycerol on mica, hence providing an unprecedented insight on interfacial friction \textit{versus} liquid's bulk dynamics. We not only evidence a massive increase of the slip length with decreasing temperature, but more unexpectedly \MLR{a non-monotonous dependency of the friction coefficient on the liquid's relaxation rate.} 

In the low molecular frequency ($f_{\alpha}\rightarrow 0$) limit, we show that liquid friction decreases \MLR{as $1/f_{\alpha}$, in good agreement with theoretical predictions for a frozen wall. On the other hand,  above the threshold frequency $f_c \sim 34$ MHz}, $\lambda$ is found to scale linearly with  $f_\alpha$. We propose that a fluctuation-induced dissipation associated with the solid internal modes adds up to the corrugated-potential contribution. \MLR{As liquid modes are blue shifted upon heating, they couple more efficiently with mica's phonons, which eventually overthrow the static corrugation contribution. Our theoretical analysis reproduces nicely the crossover between these static and dynamic frictional regimes, although a completely quantitative prediction is out of reach at this point. This picture echoes several recent works on friction and heat transfer \cite{kavokine_fluctuation-induced_2022,lizee_strong_2023,yu2023electron,noh2022effect,ma2015water}, demonstrating that electronic and phononic excitations play a key role at liquid interfaces.} Interestingly, internal wall dynamics are often omitted in molecular dynamics studies, our results show that they should in fact be given special attention \cite{herrero_fast_2020,lafon_giant_2022}.  

On experimental grounds, the dynamical friction we demonstrate here offers a new tuning knob for liquid-friction control. In recent years, the careful engineering of phonon and electron band structures has become instrumental in the design of new thermoelectric materials. This could readily be applied to the specific engineering of liquid-solid interface properties, {\it e.g.} friction, thermal conduction or even mechano-electric energy conversion.

\section*{Acknowledgement}
The authors acknowledge support by ERC project n-Aqua. B. C. acknowledges funding from a J.-P. Aguilar grant of the CFM Foundation. The authors thank N. Kavokine for many fruitful discussions. \newline

\bibliography{slippage}
\end{document}